\documentclass[preprint,12pt]{elsarticle}
\journal{Astroparticle Physics}

\usepackage{graphicx}
\usepackage{amsmath}
\usepackage{amssymb}

\begin{document}
\begin{frontmatter}

\title{The IceCube Realtime Alert System}
\emailauthor{analysis@icecube.wisc.edu}{IceCube Analysis}


\author[Adelaide]{M.~G.~Aartsen}
\author[Zeuthen]{M.~Ackermann}
\author[Christchurch]{J.~Adams}
\author[BrusselsLibre]{J.~A.~Aguilar}
\author[MadisonPAC]{M.~Ahlers}
\author[StockholmOKC]{M.~Ahrens}
\author[Erlangen]{D.~Altmann}
\author[Marquette]{K.~Andeen}
\author[PennPhys]{T.~Anderson}
\author[BrusselsLibre]{I.~Ansseau}
% ...
\address[Adelaide]{Department of Physics, University of Adelaide, Adelaide, 5005, Australia}
% ...
\address[BrusselsLibre]{Universit\'e Libre de Bruxelles, Science Faculty CP230, B-1050 Brussels, Belgium}
% ...
\address[Christchurch]{Dept.~of Physics and Astronomy, University of Canterbury, Private Bag 4800, Christchurch, New Zealand}
% ...
\address[Erlangen]{Erlangen Centre for Astroparticle Physics, Friedrich-Alexander-Universit\"at Erlangen-N\"urnberg, D-91058 Erlangen, Germany}
% ...
\address[MadisonPAC]{Dept.~of Physics and Wisconsin IceCube Particle Astrophysics Center, University of Wisconsin, Madison, WI 53706, USA}
% ...
\address[Marquette]{Department of Physics, Marquette University, Milwaukee, WI, 53201, USA}
% ...
\address[StockholmOKC]{Oskar Klein Centre and Dept.~of Physics, Stockholm University, SE-10691 Stockholm, Sweden}
% ...
\address[PennPhys]{Dept.~of Physics, Pennsylvania State University, University Park, PA 16802, USA}
% ...
\address[Zeuthen]{DESY, D-15735 Zeuthen, Germany}
\fntext[Tokyofn]{Earthquake Research Institute, University of Tokyo, Bunkyo, Tokyo 113-0032, Japan}


\begin{abstract}

Although high-energy astrophysical neutrinos were discovered in 2013, their origin is still unknown.
Aiming for the identification of an electromagnetic counterpart of a rapidly fading source,
we have implemented a realtime analysis framework for the IceCube neutrino observatory.
Several analyses selecting neutrinos of astrophysical origin
are now operating in realtime at the detector site
in Antarctica and are producing alerts for the
community to enable rapid follow-up observations.  The goal of these observations is to locate
the astrophysical objects responsible for these neutrino signals.
This paper highlights the infrastructure in place both at the South Pole site
and at IceCube facilities in the north that have enabled this
fast follow-up program to be implemented.  Additionally, this paper
presents the first realtime analyses to be activated within this framework,
highlights their sensitivities to astrophysical neutrinos and background event rates,
and presents an outlook for future discoveries.

\end{abstract}


\end{frontmatter}

\section{Introduction}
\label{sec:introduction}

Multimessenger astronomy, the combination of observations in cosmic rays, neutrinos,
photons of all wavelengths, and gravitational waves, represents a powerful tool to study the
physical processes driving the non-thermal universe.
Neutrinos play an important role in this emerging field.  The detection of
a diffuse flux of astrophysical neutrinos by IceCube~\cite{Aartsen:2015rwa, Aartsen:2016xlq} with no clearly
identified sources further motivates a multimessenger approach.
Unlike their counterparts in photons and charged cosmic rays, neutrinos' low cross section and absence of electric charge allow them
to travel the cosmological distances necessary to reach Earth from source regions without absorption or deflection.
Observation of these astrophysical neutrinos
can provide critical directional information that can be used to direct follow-up
observations.  Additionally, the detection of neutrinos from a source
is a telltale sign of high-energy hadronic interactions.  This feature
could lead to the elucidation of the accelerating mechanism which produces the most
energetic particles observed in the Universe, the highest energy cosmic rays~\cite{Agashe:2014kda}.
Several models~\cite{Waxman:1997ti,Razzaque:2005bh,1979ApJ...232..106E}
predict emission from flaring objects or other transient phenomena, requiring a
rapid start for follow-up observations to be successful.  This paper presents the IceCube
realtime system, which enables rapid identification of neutrino candidates and issues notifications
to follow-up observatories.

The IceCube neutrino detector~\cite{Achterberg:2006md,Aartsen:2016inprep} consists of 86 strings,
each instrumented with 60 digital optical modules (DOMs) spaced up to 17\,m apart over a
total vertical length of one kilometer. Strings are arranged in a hexagonal
pattern with 125\,m average horizontal spacing between neighboring strings.
Eight of the deployed strings fill in a central volume between standard strings and create the more densely
instrumented DeepCore region~\cite{2012APh....35..615A}.
The deepest modules are located 2.45\,km below the surface
so that the instrument is shielded from the large background of cosmic rays
at the surface by approximately 2\,km of ice.
Each DOM consists of a glass pressure housing containing
the photomultiplier and electronics that independently digitize the signals using
onboard electronics. The total instrumented detector
volume is a cubic kilometer of highly transparent~\cite{Aartsen:2013rt} Antarctic ice.

IceCube does not directly observe neutrinos, but rather the Cherenkov emissions from
secondary charged particles produced through electromagnetic interactions as
these secondary particles travel through the Antarctic glacial ice.  Therefore the ability
to reconstruct accurately the direction of an event recorded in IceCube is highly
dependent on the ability to reconstruct these secondary particles.

These secondary particles can produce two distinct
classes of signals within IceCube.  Track events are produced by muons, arising mainly
from the charged current interaction of muon-type neutrinos, which produce $\mathcal{O}$(km) long
light emission regions as they transit the detector.  These tracks can be reconstructed
with a directional uncertainty less than $1^\circ$, but with large energy uncertainty since an
unknown fraction of their energy is deposited outside the instrumented volume.
Shower events are produced by the
charged current interaction of electron and tau-type neutrinos and by neutral
current interactions of all neutrino types.  Shower events tend to deposit all
their energy within $\mathcal{O}$(10m), producing
a relatively isotropic deposition of light emission.  These types of events tend to have
good energy resolution ($\delta E/E\sim$15\%~\cite{Aartsen:2013vja}), but have limited angular reconstruction
in ice, with typical angular resolutions on the order of $10-15^\circ$~\cite{Aartsen:2014gkd}.

The depth of the detector and its size result in a trigger rate of approximately 2.7 kHz for penetrating
muons produced by interactions of cosmic rays in the atmosphere above the detector.
The neutrino detection rate (a few mHz) is dominated by neutrinos produced in the Earth's
atmosphere.  This large down-going background in the southern hemisphere necessitates a
higher threshold for neutrino detection relative to the earth-shielded northern hemisphere.
The first challenge of the realtime alert system is to select a
sufficiently pure sample of neutrinos, while the second is to identify the small
fraction of neutrinos that are likely to be astrophysical in origin.

IceCube has sensitivity to astrophysical neutrinos from the entire sky (4$\pi$ steradians) and operates with a high duty
factor ($>$99\%), putting it in a unique position to act as a
trigger for other observatories around the globe.
Given the limited field of view of most follow-up instruments, events from the track event class are preferred
for realtime follow-up alerts.  The smaller directional uncertainties of track events also help to limit
coincidental discoveries when sensitive telescopes are pointed at  unexplored regions of the sky.

IceCube has long had an active follow-up program.  For
several years, alerts have been sent out to optical, gamma-ray, and X-ray telescopes~\cite{GFUProceedings},
which has led to a number of interesting results~\cite{OFUPaper,PTF12csy,Evans:2015qia,GFUpaper,Santander:2016bvv}
as well as fostering a rich collaboration between electromagnetic and neutrino observatories.
These long-running follow-up programs are supplemented by
new neutrino selections that target single events deemed likely to be of astrophysical origin
in the IceCube realtime alert system.

This paper describes the technical infrastructure ($\S$\ref{sec:infra})
now in place at the South Pole and at IceCube's computing facilities in the northern
hemisphere, as well as the practical challenges of working with a remote
detector to support the realtime alert system.  It highlights the existing follow-up programs
that are now in operation ($\S$\ref{sec:alerts}) in this realtime alert framework and issuing alerts for astrophysical
neutrino candidates to follow-up observatories.  These alerts, such as~\cite{GCN}, have received prompt
observations by many observatories across the electromagnetic spectrum.

\section{Follow-up Infrastucture}
\label{sec:infra}

\subsection{Processing at the South Pole}
\label{subsec:polereprocessing}

Due to the remote Antarctic location, IceCube has established a set of automated
data collection and filtering systems that process all data received
from its DOMs.  These systems are responsible for
collecting correlated data records from DOMs, known as an event, application of
calibration information, processing waveform data from the DOMs to reconstruct
tracks and shower events, and application of event selections to select
events for consideration in neutrino search algorithms.  These systems run continuously,
processing data from the DOMs as rapidly as possible on a dedicated
computing cluster located at the detector site.  These systems also
host event selections used to identify astrophysical neutrino events, generate
alert messages, monitor the health of the IceCube detector, and transmit this
information north for dissemination to the astronomical community with minimal delay.

The IceCube data acquisition (DAQ) system is responsible for managing communication with and control
of all deployed DOMs~\cite{Abbasi:2008aa,Abbasi:2010vc}.  Time-stamped signals from all DOMs
are received by the DAQ~\cite{Aartsen:2016inprep}.  The primary trigger for
neutrino alerts searches for 8 DOMs receiving photon signals
in a $5\mu s$ time window.  Once the trigger has been satisfied, the DAQ records
all DOM signals from a +6/-4 $\mu s$ window around the trigger time into a single event.  The
recorded events from each $\sim$1 second slice of data are immediately made available to the online
processing and filtering system.

The online processing and filtering system~\cite{Aartsen:2016inprep} distributes
each event to one member of a farm of $\sim$400 identical, dedicated calibration
and filtering client processors.   These clients include software for calibration of DOM digitized waveforms and
extraction of light arrival times from the recorded DOM waveforms.  The photon arrival and amplitude information
is extracted from the calibrated waveforms using the DOM response to single photons using a non-negative
linear least squares algorithm~\cite{Aartsen:2013vja}.  The relative timing
of the DOM signals is calibrated by the DAQ to UTC times with a measured accuracy of
1.2~ns~\cite{Aartsen:2016inprep}.  The calibration values used online are identical to those
used in offline analyses, and show little year-to-year variation.

This system also includes several reconstruction algorithms that characterize
each event's extracted light arrival information against the expected patterns from
track and shower events to determine the direction, position and energy of each event~\cite{Aartsen:2013vja}.
Based on these reconstructions, approximately 1\% of these events are selected to
be potentially of neutrino origin and are processed with additional,
more sophisticated and computationally intensive reconstructions (the "OnlineL2" selection).
The selection targets well-reconstructed, track-like events with a charge threshold that depends
on the reconstructed track direction, with more stringent cuts applied in the atmospheric
muon dominated down-going region~\cite{GFUpaper}.  The OnlineL2 selection has been in operation
since 2011 with each year seeing incremental improvements that bring better tools developed
offline to the online system.   For the gamma-ray, optical, and x-ray follow-up program (see $\S$\ref{subsec:ofu_gfu_alert})
the OnlineL2 selection serves as the pre-selection for these online neutrino searches.

The online processing and filtering clients are able to select any triggered events
that pass established event quality, energy, and
topology criteria for the alert systems based on the online event reconstruction information.
These selection criteria search for single events, such as the
rare high-energy astrophysical neutrinos, are established and verified in offline studies, and
are derived from similar selections used in published
analyses~\cite{Aartsen:2014cva,FirstPeVFound,Aartsen:2013dsm,HESEScience}.  These selections
include a high-energy starting event selection (see $\S$\ref{subsec:hese_alert}) and
an extremely high-energy track selection (see $\S$\ref{subsec:ehealert}).

The results from this client farm are returned to the online processing and filtering system,
where data files containing all events are created and archived.  Each event is processed in
the order received, maintaining a strict first-in, first-out order.  This  architecture
puts practical limits on the complexity of reconstructions performed in the filtering client
as all events must be reconstructed within $\sim$30 seconds to prevent pileup.
Several computationally complex reconstructions performed offline~\cite{Aartsen:2013vja}
can require minutes to hours to evaluate a single event and are not
supportable by the limited computational power available at the South Pole.
Events passing the online
alert system criteria are forwarded to a dedicated online alert system.
Typically, alert information is presented to the online alert
system with a delay of about $\sim$20 seconds.

The online alert system receives events selected for alert generation and immediately
creates messages for transmission to the northern hemisphere data center.
The first message contains a short JSON-formatted\footnote{\texttt{http://www.json.org/}}
message containing the critical information needed for inclusion in automatic
alerts to astronomical partners (event time, run information, direction and energy).  A second, larger
message is generated encoding a compact data record of the extracted light
arrival information from the DOMs.  This second data record is also transmitted north in order to
start more computationally intensive follow-up reconstructions and alert quality
verification checks.   This splitting of alert information across multiple messages is done
to avoid transmission delays in the critical information that is used in the initial automatic alerts.

To ensure IceCube is operating in a stable manner before notifying follow-up instruments,
the DAQ and the online processing and filtering system
track several detector health monitoring quantities.  These include the rate of triggered events,
rate of filtered events, number of active DOMs, and several other detector environmental
criteria that have been found to be good indicators of detector health and data quality.

\subsection{Data Transfer}
\label{subsec:datatransfer}

Alert messages from the online alert system and all detector health information is reported
to the detector experiment control system, IceCube Live (I3Live)~\cite{Aartsen:2016inprep}.
The I3Live system has several data transport methods available to move information from
Antarctica to the primary data center at the University of Wisconsin, Madison, each with different
bandwidth, message size limits, and latency.

\begin{figure}[!ht]
\centering
\includegraphics[width=0.7\textwidth]{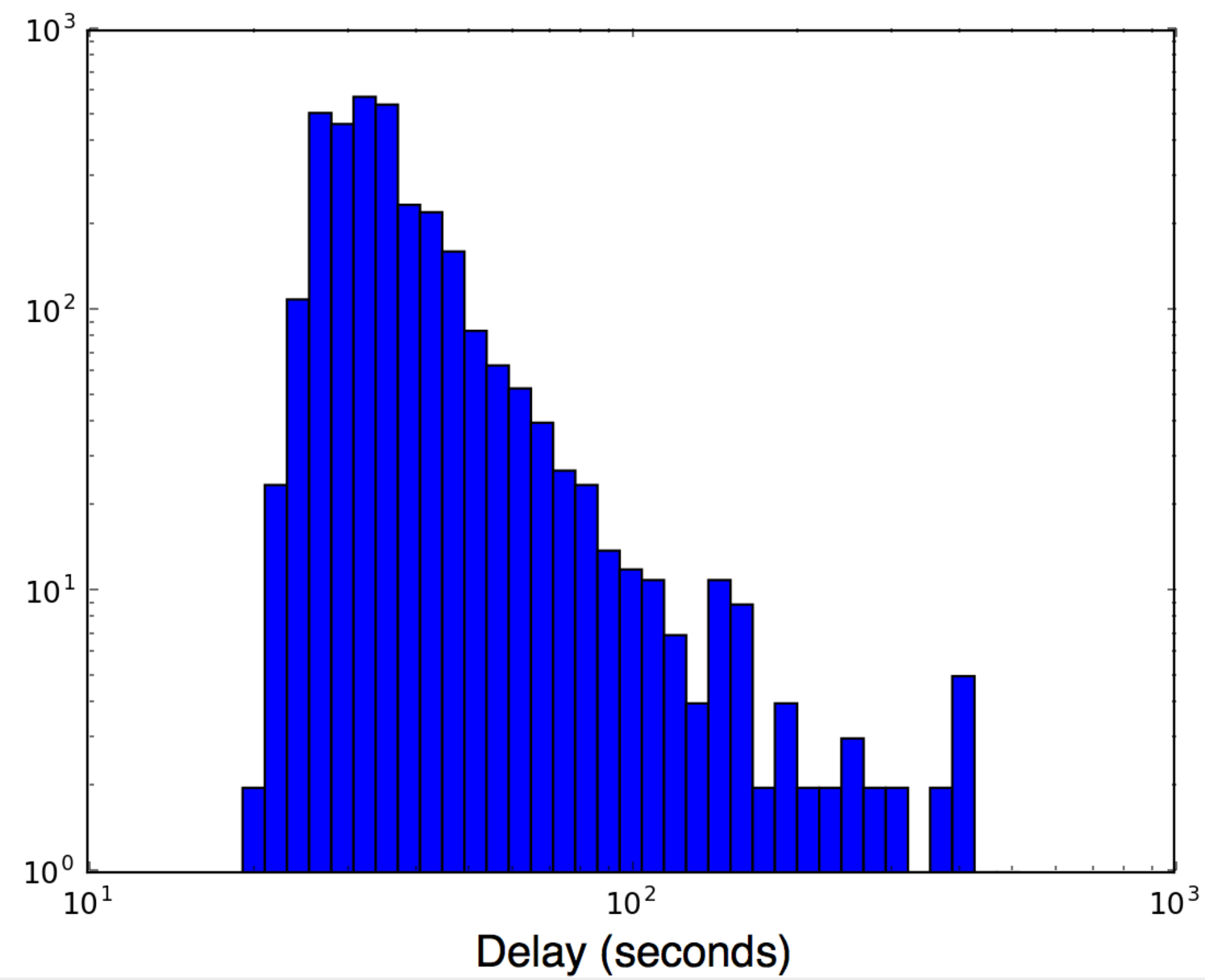}
\caption{Histogram of total alert message latency for the Iridium
RUDICS data messaging system, measured from time the event triggers the data
acquisition system at the South
Pole until the alert is received in the northern hemisphere data center.
The median message latency is 33~s.}
\label{fig:infra_rudics}
\end{figure}

JSON messages containing the alert summary
information, as well as messages containing compact records of light arrival
information, are transferred by the Iridium RUDICS
system\footnote{\texttt{https://www.iridium.com/}}.  The IceCube detector
utilizes 2 links, each with 2.4~kbps bandwidth.  Messages
are sent north without message size restrictions, but can experience some delays due
to higher priority data traffic.  Total message latencies, including the $\sim$20 second
event processing and filtering time, are shown in Fig.~\ref{fig:infra_rudics}
for the short JSON alert messages, with a median total message latency of 33~s.

\subsection{Alert Generation}
\label{subsec:alertgen}

Alert messages arriving from South Pole via I3Live are immediately stored in a
dedicated database and distributed to a set of follow-up analysis clients.
This distribution employs a ZMQ\footnote{\texttt{http://zeromq.org}} framework
with a publisher-subscriber model, where the I3Live acts as a publisher,
distributing the alert information to all analysis clients
which have subscribed to a particular type of alert message.
This setup provides a scalable and stable platform for
distributing event streams of varying rates to a multitude
of clients, while decoupling the operation of subscribers from each other
and the publisher, allowing each subscriber to operate in an independent manner.

The follow-up clients use a shared library, which provides methods to communicate
with the event publisher, assess the detector status, do basic analysis tasks
and generate properly formatted alert messages to other observatories using
several forms of automated communication, such as email or
VOEvent~\cite{Seaman:2011cb} messages.

Each realtime analysis has a dedicated follow-up client process that is triggered by
the arrival of an incoming event and determines whether an alert should be generated.  These
follow-up clients have the ability to fetch previously recorded events from the database for correlation studies,
as well as to monitor quantities vital to the operation of the detector to ensure
that certain data quality criteria are met. Depending on the observation plan established with the follow-up observatory,
the alert generation can be inhibited, if the source is not visible to the observatory
at the time of the alert or during the following nights.

\subsubsection{Detector Stability Monitoring}
\label{sec:OnlineMonitoring}

All event selections and alert mechanisms depend on events being acquired in
well-determined and well-understood operational states of the detector.
For traditional offline analyses, periods of data taking are split in segments with a duration of eight hours,
then each segment is manually inspected to ensure good data quality.
Data quality can be impacted by several operational issues, such as groups of DOMs or entire strings failing to
deliver data, or by operation of light-generating calibration devices within the detector.

\begin{figure}[t!]
 \centering
 \includegraphics{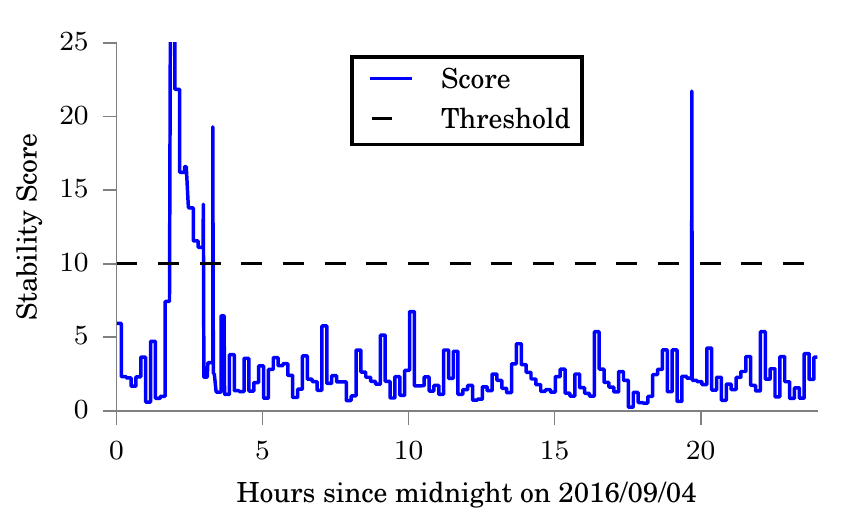}
 \caption{Output of the online stability monitoring from an anomalous day.
          The stability score encodes information about the quality of the online data taking.
          In this example a faulty power supply (at hour $\sim$2) and a full restart of the detector (at hour $\sim$19)
          cause some downtime, where the stability score exceeds the threshold of 10, during otherwise stable operations.
					The bad intervals are correctly identified.}
 \label{fig:stability_score}
\end{figure}

To ensure a quick alert generation,
realtime alerts cannot rely on manual determination of data quality.   Therefore an automated system to
monitor the detector stability continuously has been implemented~\cite{GFUpaper}.
There are three ingredients,
which are directly related to the different stages of the event selection (see $\S$\ref{subsec:polereprocessing}):
\begin{itemize}
	\item The rate of primary DAQ triggers.
	\item The rate of the well-defined muon track events selected by the online filter system.
	\item The rate of the events selected by the OnlineL2 selection.
\end{itemize}
These quantities are recorded in ten minute intervals.
An exponentially weighted moving average is formed over past measurements and the rates in each
new bin are compared to this average
(weighted with their statistical uncertainty).
The deviation from the long term average yields a stability score
from which the goodness of the data can be measured.
An example of one day of mostly stable conditions with two outages in between
is given in Fig.~\ref{fig:stability_score}.

The difference in good quality data taking periods measured by the online criteria and the offline monitoring system is less
than one percent, with the online monitoring rejecting slightly more time periods out of an abundance of caution.

\subsubsection{Alerts, Revisions and Retractions}
\label{sec:AlertsAndRevs}

Alerts are issued to the observational community in several ways, both through public and private
channels.  Public alerts utilize the Astrophysical Multimessenger Observatory Network (AMON) system~\cite{AMONpaper}
as a gateway to the Gamma-ray Coordinates Network (GCN)~\cite{gcnweb} and are immediately available to follow-up observatories.
Each alert has a well-defined structure and content, which is published online\footnote{\texttt{http://gcn.gsfc.nasa.gov/amon.html}}.    Private
alert communication is generally done via electronic mail messages directly to observatory operation centers or by private
distribution lists via AMON and GCN.

The event reconstruction information can be refined within a few hours of an initial alert with better
angular reconstruction and improved distinction between track-like and shower-like events.
In the initial alert, directional reconstructions from the OnlineL2 selection performed at the South Pole are used.
Once the full event information arrives in the northern data center, additional reconstructions can begin on larger computer clusters.
These CPU-intensive reconstructions evaluate the likelihood of different arrival directions using a series of ever-finer directional grids on the sky to
determine the best-fit direction.
Angular resolution estimates at different confidence levels are estimated using Wilks's theorem that has been calibrated using
known angular errors measured with simulated neutrino data samples.

These refined reconstructions improve angular resolution by more than 50\% (to $\le 1^\circ$ for most tracks) and  also provide information on the
energy loss within the detector. The energy loss profile, in conjunction with the likelihood direction scans,
provide important inputs to further distinguish between track-like and shower-like events.  The angular resolution for shower-like events,
typically $\sim$10-15$^\circ$, is too large for follow-up with most telescopes.

Once these refined reconstructions are complete, a revision to the original alert can be created reflecting the updated event information.
The revised alert retains the same event number information as the original alert, but the revision number is increased by 1.
Additionally, any alert that is determined to originate from misreconstructed background events or other instrumental effects will be retracted
within hours following the original alert.  Refined reconstructions for all high-energy starting
events (see $\S$\ref{subsec:hese_alert}) and extremely high-energy track events (see $\S$\ref{subsec:ehealert})
are automatically performed and are accompanied by alert revisions upon completion.  Refined reconstructions
are available for other events on an as-needed basis.

\section{Realtime Alert Systems}
\label{sec:alerts}

IceCube presently operates several alert systems utilizing the realtime framework described in the
preceding section.  The X-ray, optical, and gamma-ray follow-up program described in
$\S$\ref{subsec:ofu_gfu_alert} has been in operation for several years and is accompanied by two
new online streams beginning in 2016.  The first new alert system selects track-like high-energy starting
events (HESE, $\S$\ref{subsec:hese_alert}) and the second alert system targets extremely high-energy through-going
tracks (EHE, $\S$\ref{subsec:ehealert}).

The infrastructure in place allows for independent, simultaneous operation of alert systems that search
for different signals.   The HESE and EHE alert systems both trigger on the detection
of single events, while other follow-up programs are triggered by the accumulation of neutrino candidates
consistent with coming from a single point in the sky.
As new analysis techniques are developed to quickly isolate astrophysical neutrino candidates,
they will be moved to the realtime alert system to generate
triggers for interested follow-up observatories.

\subsection{Gamma-Ray, Optical and X-Ray Follow-up}
\label{subsec:ofu_gfu_alert}
The gamma-ray, optical and X-ray follow-up program is designed to detect bursts of several neutrino-like events that,
when considered alone, would not be distinguishable from background. The main background for up-going events, i.e., events
coming from the northern hemisphere, are atmospheric neutrinos. These neutrinos arise from decays of charged pions and kaons that are created
by cosmic rays striking the atmosphere. For down-going events the background is dominated by lower energy muons from these atmospheric showers, so
only high-energy tracks are selected~\cite{Aartsen:2014cva}.
These backgrounds are well understood, with well measured rates and isotropic angular distributions in any selected zenith range.
This follow-up program searches for statistically significant clustering in time and space of the observed neutrino candidates,
and uses any spatial and time correlation as an indication of a potential neutrino burst.  Given the low expected rate of
alerts from true neutrino bursts, alert thresholds are set to generate follow-up alerts for a
few background over-fluctuations per year.  A common event selection, described in $\S$\ref{sec:gfufilter} is used
in two different neutrino burst time scale searches, optical and X-ray follow-up (OFU, $\S$\ref{sec:ofutrigger}),
searching time scales up to 100 s,
and the gamma-ray follow-up (GFU, $\S$\ref{sec:gfutrigger}),
searching time scales up to 3 weeks.

\subsubsection{Event Selection}
\label{sec:gfufilter}

\begin{figure}[t]
 \centering
 \includegraphics{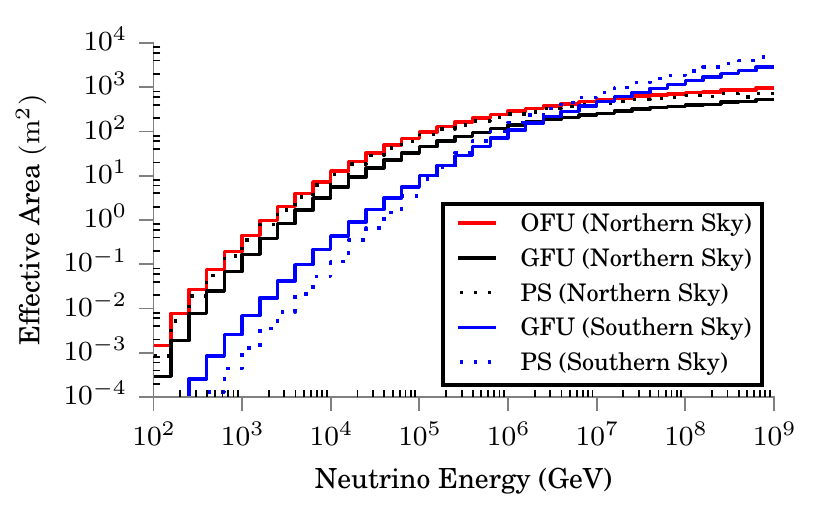}
 \includegraphics{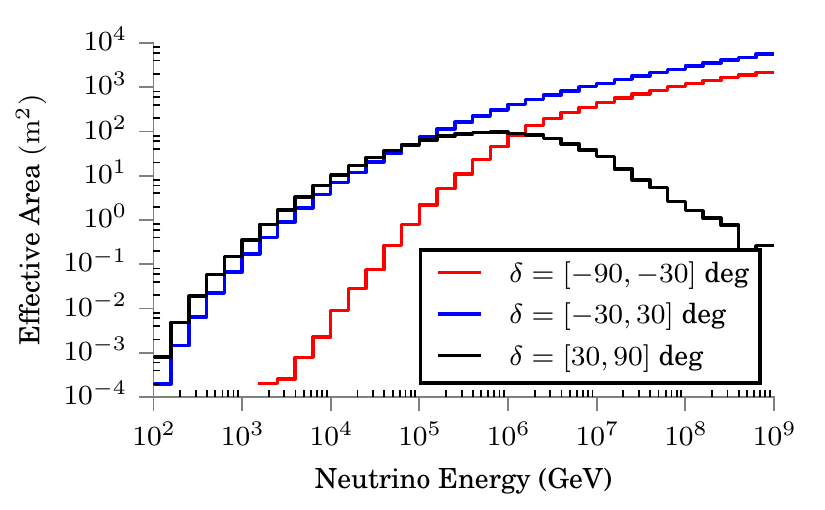}
 \caption{Effective areas for different subsets of the online neutrino event selection.
 The upper panel compares the OFU and GFU subsamples to each other, as well
 as to the published point source search selections~\cite{Aartsen:2016oji},
 and the lower panel compares different declination bands within the GFU subsample.
 The OFU subsample is only available in the northern sky,
 whereas the GFU subsample covers both hemispheres.
 }
\label{fig:eff_area_gfu_ofu}
\end{figure}

Both OFU and GFU are based on the same neutrino event selection.  This selection starts with
the OnlineL2 pre-selection ($\S$\ref{subsec:polereprocessing}) which
selects tracks that are potentially neutrino generated and contains results from more sophisticated track and energy reconstructions
as well as enhanced angular uncertainty estimators~\cite{Aartsen:2014cva}. Prior to May 2016, the OFU and GFU selections used
independent event selections that yielded similar event samples, but they have since been unified into the single
selection described here.
The results of these reconstructions are used as the input to a multivariate classifier.
Further reduction of the atmospheric muon background and separation of an
astrophysical signal is achieved by training boosted decision trees
(BDTs)~\cite{Freund97adecision-theoretic}.
The training is done separately for the northern and southern hemisphere to account
for the different kinds of background encountered in each region,
yielding a final event selection rate of 5~mHz,
equally divided between both hemispheres.
A description of the variables entering the BDTs can be found in~\cite{GFUpaper}.

\begin{figure}[t]
 \centering
 \includegraphics{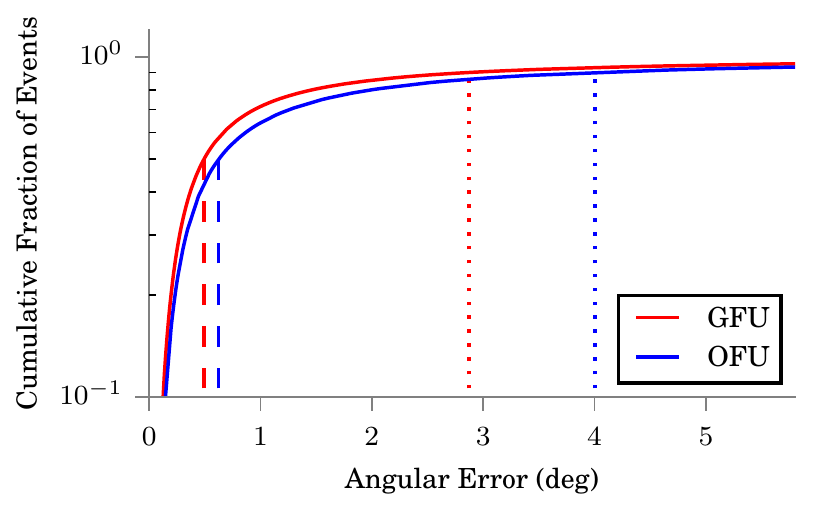}
 \includegraphics{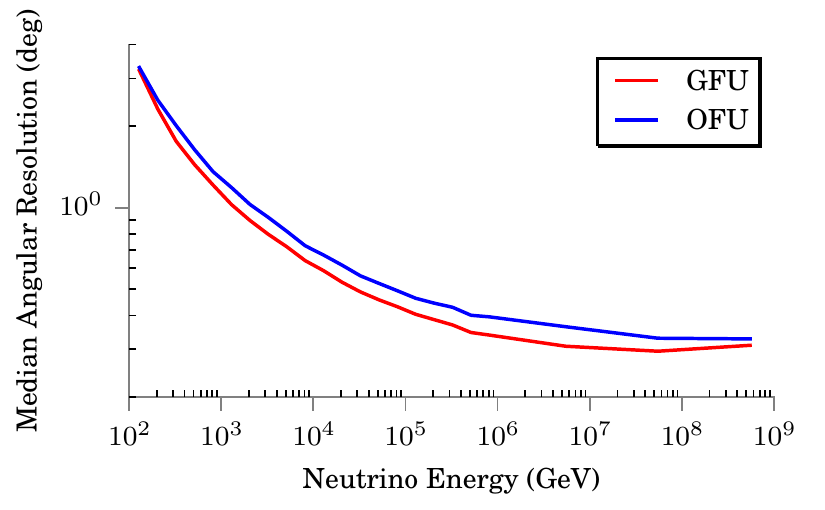}
 \caption{Angular error of events in the GFU and OFU analyses.
The angular error is the opening angle between the online reconstructed
direction and the true direction of simulated neutrinos.
The upper plot shows the cumulative distribution,
with the dashed (dotted) line highlighting the 50\% (90\%) containment.
The lower plot shows the median angular error as a function of the
true neutrino energy.}
\label{fig:ang_err_gfu_ofu}
\end{figure}

Two different subsamples are defined from the output of the BDTs:
\begin{itemize}
\item The OFU subsample, limited to up-going tracks from the northern sky, uses a relaxed BDT cut,
with an event rate in this hemisphere that is slightly higher (3 mHz) than for the GFU analysis.
\item The GFU subsample selects tracks from the entire sky.
In the northern sky, the BDT cut is more selective than that of the OFU analysis, leading to a lower event rate (2 mHz) in this region.
The BDT cut in the southern sky is chosen such that the event
rate is constant in all declinations and is matched to the northern sky.
\end{itemize}

The different BDT cuts are selected to achieve the best sensitivity to neutrinos
for the different time scales searched relative to the expected backgrounds.  A smaller
search time window is used in the OFU analysis to reduce background events and to allow for the more
inclusive event selection to be used.
The effective areas for neutrinos, as used in the OFU and GFU selections, are
shown in Fig.~\ref{fig:eff_area_gfu_ofu}.  The angular error for selected neutrino events, and its dependence
on the energy of the neutrino, is shown in Fig.~\ref{fig:ang_err_gfu_ofu}.
All events selected by the BDT for either OFU or GFU are immediately transferred
from the South Pole to the north, where they are made available to the OFU and GFU analysis clients as described
below.

\subsubsection{OFU Alerts}
\label{sec:ofutrigger}
The optical and X-ray follow-up~\cite{OFUPaper,PTF12csy,Evans:2015qia}
aims for the real-time detection of neutrino bursts on time scales of up to $100$\,s,
which are predicted to be produced in gamma-ray bursts or supernovae with choked jets~\cite{Razzaque:2005bh,Tamborra:2015fzv,Senno:2015tsn}.
Given the reduced background rate from the Earth's shielding of cosmic rays, the OFU
program focuses on alerts from the northern hemisphere (see Fig.~\ref{fig:eff_area_gfu_ofu}.)
The program has been operating since December 2008 when interesting neutrino events were
initially  forwarded to the now-decommissioned ROTSE telescopes~\cite{2003PASP..115..132A}.  Starting in August 2010,
the Palomar Transient Factory~\cite{2009PASP..121.1395L} started receiving OFU alerts, and in February 2011
the inclusion of the Swift-XRT~\cite{Burrows2005}, targeting GRB afterglows, marked the expansion of the program
to the X-ray regime. In 2016, the program was again expanded in its optical capabilities, enabling more complete
sky coverage by
distributing alerts also to the MASTER telescopes~\cite{2010AdAst2010E..30L}.  An extension to
ASAS-SN\cite{asassnweb} and LCO~\cite{2014SPIE.9149E..12P} telescopes is being planned.

OFU selects neutrino multiplets, requiring at least two events within $100$\,s and with an angular difference of less
than $3.5^\circ$. Multiplets with a multiplicity larger than two are immediately forwarded to optical and X-ray telescopes.
For doublets an additional quality cut is applied. This quality cut parameter, $\lambda$, is defined as follows:

\begin{align}
    \lambda = \frac{\Delta\Psi^2}{\sigma_q^2} + 2 \ln(2 \pi \sigma_q^2) - 2 \ln\left( 1 - \exp\left(-\frac{\theta_A^2}{2\sigma_w^2}\right) \right)
            + 2 \ln\left( \frac{\Delta T}{100\,s} \right)
  \label{eq:alert-llh}
\end{align}
where the time between the neutrinos in the doublet is denoted as $\Delta T$,
and their angular separation as $\Delta\Psi$. The quantities $\sigma_q^2 =
\sigma_1^2 + \sigma_2^2$ and $\sigma_w^2 = \left(1/\sigma_1^2 +
1/\sigma_2^2\right)^{-1}$ depend on the per-event estimated directional
uncertainties $\sigma_1$ and $\sigma_2$ of the two neutrino events, typically
$\sim 1^\circ$.
The angle $\theta_A$ corresponds to the circularized angular radius of the
field of view (FoV) of the follow-up telescope. The quality parameter ($\lambda$) is smaller for more signal-like alerts, which have small
separation $\Delta\Psi$, small time difference $\Delta T$ and a high
chance to lie in the FoV of the telescope. Thus, $\lambda$ is a useful
parameter to identify signal-like doublets, and reduce the rate of
background alerts. For each follow-up
instrument, a specific cut on $\lambda$ is applied in order not to exceed the granted number of alerts per telescope and to send the
most significant alerts to the follow-up instruments, as well as ensuring that any potential alert position
is available for observation (e.g., at a sufficient distance from the Sun and Moon).  The circularized angular radius
($\theta_A$) and maximum allowed alerts per year determine the $\lambda$ cut level for each follow-up telescope.
PTF and Master ($\theta_A = 0.9$) receive up to 7 alerts per year, while the
Swift-XRT ($\theta_A = 0.5$) receives up to 3.  Given the small angular aperture, Swift-XRT observations are done by tiled
observations about the alert position.
Longer term observations are scheduled as required based on the results
of the initial observation.


The combined direction of the events in the multiplet and multiplet detection date and time
are sent to the telescopes as an alert. The combined direction is the weighted
arithmetic mean, weighting the individual directions with their inverse squared error, given by the per-event directional
uncertainty $\sigma_i$. The error of the combined direction is given by $\sigma_w = (\sum_{i=1}^{N} (1/\sigma_i)^2)^{-1}$, where $N$
is the multiplicity of the alert.

\subsubsection{GFU Alerts}
\label{sec:gfutrigger}

The gamma-ray follow-up (GFU) searches for neutrino bursts
on time scales of up to three weeks, tailored to the variability observed in several sources by imaging atmospheric
Cherenkov telescopes.  The GFU program is described in detail elsewhere~\cite{GFUpaper}.
The GFU analysis searches for  an excess of neutrino events
in the vicinity of sources from a predefined source catalog.
The list of monitored sources is based on the second Fermi point-source catalog~\cite{2012FGL},
containing mostly BL-Lac objects and FSRQs,
which have exhibited previous variable behavior and are visible to the follow-up
telescopes.  When a  significant cluster is observed, the information is forwarded to
the MAGIC~\cite{MAGICPerformance}
and VERITAS telescopes~\cite{VERITASPerformance}
(depending on source visibility at each site and distance from the Moon) to search for a coincident flare in very high-energy gamma rays.
The program has been operating since 2012 with an online event selection covering the northern hemisphere sky.
In 2015, it was extended to include the southern hemisphere sky.

A maximum-likelihood based search for point sources~\cite{Braun:2008bg}
is paired with a time-clustering search algorithm~\cite{2011TimeClustering} to search for neutrino bursts from
a given source direction.
Starting with the last event observed, time windows of up to three weeks in the past are
tested to determine the most likely time frame of a flare,
identified as a significant deviation of the number of detected neutrinos from the expected background.
For each time window, the detector uptime
(see $\S$\ref{sec:OnlineMonitoring}) is considered in the likelihood calculation.
A detailed description of the algorithm can be found in~\cite{GFUpaper}.

In early operation, the analysis used a binned counting method and
a basic event selection which required event quality parameters to exceed a fixed threshold.
With the aforementioned upgrade to BDTs and the maximum-likelihood search,
the sensitivity has improved by 25\% at the horizon and 65\% towards the North Pole,
which supports an increase in the number of monitored sources from 109 to 184,
by requiring a higher significance threshold for the alert generation.
On average, two alerts per year are expected from background.

With the inclusion of the southern sky in the event selection, a collaboration with the HESS telescope~\cite{Schussler:2015aia} is in preparation.
Furthermore, future work will extend this program to flares arising from the entire sky on
arbitrary time scales (the upper time scale is given by the results from
the static point source search~\cite{Aartsen:2014cva}).

\subsection{HESE Alerts}
\label{subsec:hese_alert}
The IceCube high-energy starting event (HESE) search has resulted in a clear detection $(>6.5\sigma)$ of astrophysical
neutrinos~\cite{HESEScience, FirstPeVFound, ICRCKopper}.
However, the nature of the sources responsible for these neutrinos is not yet known.
The sources of these neutrinos may be identified by the detection of an electromagnetic
counterpart in rapid follow-up observations.

IceCube has detected 54 HESE neutrino candidates in 4 years of data~\cite{ICRCKopper}.
These events have interaction vertices inside the detector fiducial volume and are classified in two main categories:
track-like events from charged-current interactions of muon neutrinos (and potentially
from the $\sim$18\% of tau neutrino interactions that produce a high-energy muon)
and shower-like events from all other interactions (neutral-current interactions and charged-current interactions of electron neutrinos and most tau neutrinos).
The HESE data is dominated by shower-like events.   In the 4-year data sample, there are 14 track-like events, while the remaining 40 are shower-like events.
Given the better angular resolution of the track events,
only the track-like events identified online are considered for HESE alerts and distributed publicly via
AMON and the GCN network.

\subsubsection{HESE Track Selection and Alerts}
The astrophysical signal is most prominent at high energies where contained neutrino interactions result in
a significant amount of light in the detector, and therefore a large amount of total charge detected by the DOMs.
We use the total charge observed within 5 $\mu$s of the event start time in order to cut out light from combined low energy events.
All hits in the more densely instrumented DeepCore portion of the detector~\cite{2012APh....35..615A} and on any single DOMs containing
more than 50\% of the total charge in the event are excluded to prevent the signal from a single DOM very close to a particle track
from dominating the charge measurement.
Here, only events with $\ge 6000$ photoelectrons are considered for HESE alerts.  The primary background for HESE tracks
arises from rare atmospheric muon events that evade the veto criteria~\cite{HESEScience}.

Additionally, for an event to be considered as a HESE track, it must also exhibit signal-like and track-like characteristics.  This is
parameterized by the ``signal\_trackness'' parameter, a number between 0 and 1.
To calculate this number, Monte Carlo simulations including both signal and background events have been considered.
Signal events were simulated with an energy spectrum of E$^{-2.58}$, the best fit spectrum observed by the IceCube HESE analysis~\cite{ICRCKopper}.
A Bayesian approach has been used to calculate the probability that a HESE event is a track-like signal event:
\begin{equation}
{\rm Signal\_Trackness} = \frac{f_{\rm track} P_{\rm track}}{{f_{\rm track} P_{\rm track}+f_{\rm shower} P_{\rm shower}+ (f_{\rm bkg}/f_{\rm sig}) P_{\rm bkg}}},
\end{equation}
\noindent where $P_{\rm track}$, $P_{\rm shower}$, and $P_{\rm bkg}$ are the PDFs of log-likelihood ratios (value from the shower reconstruction divided by that of the
track reconstruction) for track-like events, shower-like events, and backgrounds, respectively.
The variables $f_{\rm track}$, $f_{\rm shower}$, $f_{\rm bkg}$, and $f_{\rm sig}$ are the prior probabilities, with $f_{\rm track} = 1 - f_{\rm shower}$
and is given by:
\begin{equation}
f_{\rm track}=\frac{R_\mu R_{\mu,\rm{cc}}+R_\tau R_{\tau,\rm{cc}} R_{\tau,\rm{cc},\mu}}{R_e+R_\mu+R_\tau}.
\end{equation}
Based on studies of simulated events, $R_e:R_\mu:R_\tau$ is $2.48:1.0:1.52$.  Also, $R_{\mu, \rm{cc}}=0.78$, $R_{\tau,\rm{cc}}=0.86$ are the fractions of $\nu_\mu$
and $\nu_\tau$ events interacting via CC, respectively, and $R_{\tau,\rm{cc},\mu}$ is related to the branching ratio of $\tau \rightarrow \mu$.
Note that this number is approximately half of the $\tau \rightarrow \mu$ branching ratio of 0.18, because only half of the $\tau \rightarrow \mu$
decays put enough energy into the muon for it to be detected as a high energy muon.
The quantity $f_{\rm bkg}/f_{\rm sig}$ is the ratio of the background to signal event rate that is dependent on charge.

Alerts are sent only for events having
signal\_trackness $\ge 0.1$. This yields about 1.1 signal-like track-like events from an E$^{-2.58}$ astrophysical
spectrum~\cite{ICRCKopper} and about 3.7 total background events per year.  Thirteen of the 14 manually identified
track events in the published HESE samples~\cite{ICRCKopper} would also be identified as tracks by this signal\_trackness
criteria.  The effective areas for this selection, for the entire sky and the northern and southern skies separately, are
shown in Figure~\ref{fig:ehe_effa}.

Events passing these criteria generate a public alert via the GCN network sent via AMON.  These alerts
contain the best-fit source direction and uncertainty from the online reconstruction, date and time of the event, total measured charge, and
signal\_trackness value.

Studies of simulated HESE track events provide estimations of HESE angular errors in real-time.
The angular separation between the true neutrino direction and online reconstructed direction is an estimate of how well our online reconstruction performs.
Events with signal\_trackness $\ge 0.1$ have a median angular error of $0.4^\circ$ to $1.6^\circ$
($1.2^\circ$ to $8.9^\circ$ for 90\% containment) based on the properties of the individual events in the simulated HESE track samples.
Events with larger track length inside the detector are better reconstructed and therefore have smaller angular errors.
Events passing a minimum reconstructed track length in the detector of 200~m (~80\% of the events) have a median angular error of $0.55^\circ$
($1.89^\circ$ for 90\% containment) as illustrated in Fig.~\ref{fig:angularres}.
Those events that do not pass a minimum track length cut (~20\%), but still pass the signal\_trackness selection,
are reported with an upper limit fixed median angular error of $1.6^\circ$ ($8.9^\circ$ for 90\% containment).
After detection and initial alert generation, follow-up reconstructions of the HESE events by computer clusters in the northern hemisphere are immediately started,
and revised directional coordinates and improved angular uncertainty are released within a few hours of the initial alert (see $\S$\ref{sec:AlertsAndRevs}).

\subsubsection{Options for Increasing the Signal-to-Noise Ratio}
\begin{figure}[t]
\centering
\includegraphics[width=0.7\textwidth]{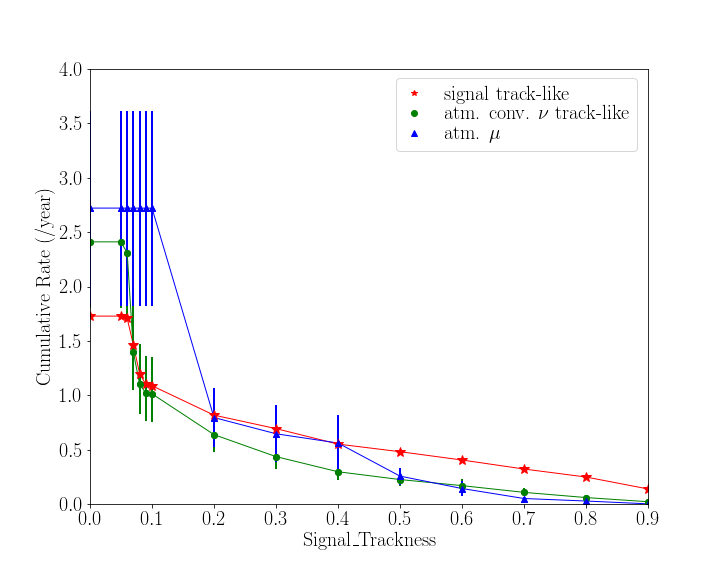}
\includegraphics[width=0.7\textwidth]{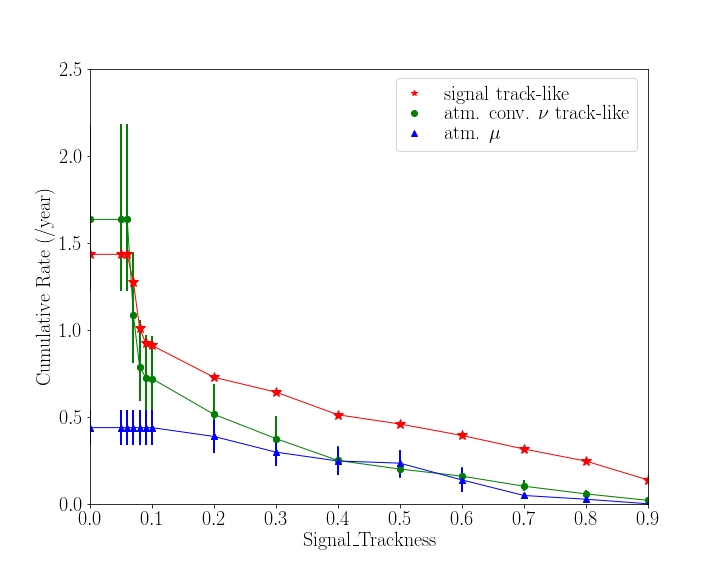}
\caption{Rate (number of events per year) vs. different cuts on the signal\_trackness estimator for HESE signal and background with
charge  $>$~6000~p.e. (upper panel) and $>$~7000~p.e. (lower panel). HESE track events are shown in red, the atmospheric conventional neutrinos in green,
and the atmospheric muons in blue. Error bars represent the standard deviation of the rates.
The muon background error bars are larger than the other two components due to
low statistics of the computationally-intensive simulation of the muon background.}
\label{fig:HESE-rate}
\end{figure}

Each HESE alert contains quantities that can be used by rapid follow-up observatories to increase the signal-to-noise ratio.
Requiring a larger signal\_trackness results in less background with some loss in signal efficiency.
Fig.~\ref{fig:HESE-rate} shows the rate versus different cuts on signal\_trackness for three different neutrino fluxes
using simulated data:
astrophysical track events in red, track-like atmospheric conventional neutrino background in green, and track-like atmospheric muon
background in blue. Contributions from a prompt neutrino flux are expected to be small~\cite{HESEScience}.  Fig.~\ref{fig:HESE-rate}
also shows the rate (number of events per year) vs. different cuts on signal\_trackness for signal and
background simulated HESE events with charge \textgreater 7000 p.e.

Another option is to require larger deposited charge.
Table~\ref{table:hesecharge} shows the signal and background rates (expected number of events per year) for different charge cuts as
well as the signal to noise ratio (SNR).
All numbers in this table are for events with signal\_trackness $\ge 0.1$.
The follow-up observatories can decide which events to observe based on their charge, signal\_trackness, and rate information.

\begin{table}
\caption{Different cuts on HESE event charges, measured in photoelectrons, result in different signal, noise rates, and signal to noise ratios.
In parentheses, the contribution from the northern and southern hemispheres are separated.  Signal events are equally distributed in the northern
and southern hemispheres, while backgrounds are stronger in the down-going southern hemisphere.}
\small
\begin{tabular}{|p{1.5cm}|p{3.6cm}|p{3.6cm}|p{3.4cm}|}
\noalign{\smallskip}
\hline
Charge & Signal Rate (yr$^{-1}$) $(R_s)$ & Background Rate (yr$^{-1}$) $(R_b)$ &  SNR$={R_s/{R_b}}$ \\
\hline
6000 & 1.09 (0.50 N + 0.59 S) & 3.73 (0.67 N + 3.06 S) & 0.29 (0.75 N, 0.19 S) \\
\hline
6500 & 1.00 (0.47 N + 0.53 S) & 2.81 (0.58 N + 2.23 S) & 0.36 (0.81 N, 0.24 S) \\
\hline
7000 & 0.91 (0.42 N + 0.49 S) & 1.16 (0.49 N + 0.67 S) & 0.78 (0.86 N, 0.73 S) \\
\hline
7500 & 0.84 (0.38 N + 0.46 S) & 0.92 (0.41 N + 0.51 S) & 0.91 (0.93 N, 0.90 S) \\
\hline
\end{tabular}
\label{table:hesecharge}
\end{table}


\subsection{EHE Alerts}
\label{subsec:ehealert}

The extremely-high-energy (EHE) neutrino alert stream is based on an offline
search for GZK, or cosmogenic, neutrinos that resulted in the discovery of
the first observed PeV-scale neutrinos~\cite{FirstPeVFound}.
The analysis selection is simple and robust, making it a natural candidate
to move into the online alert framework where computing resources are limited.

The offline diffuse EHE analysis targets neutrinos with energies
of  $\sim$~10~PeV to 1~EeV, where the expected event rate in the most
optimistic case is $\sim$1 event per year~\cite{EHEPaper2}.
When moving this analysis into the realtime
framework, the event selection was modified in order to increase the sensitivity
to astrophysical neutrinos, specifically with neutrino energies in the 500~TeV
to 10~PeV range, targeting track-like events, which have good
angular resolution ($<1^{\circ}$).


\subsubsection{EHE Event Selection and Alerts}
\label{subsubsec:eheselection}

The EHE alert selection requires a minimum deposited charge of $10^{3.6}$ photoelectrons  (NPE) detected in
DOMs in the detector volume as well as at least 300 DOMs
registering a signal.  The events are then fit with a track hypothesis and the
fit quality parameter (charge weighted $\chi^{2}$)~\cite{2014NIMPA.736..143A} is required to be consistent
with well reconstructed tracks.

Due to large background contamination from atmospheric muons originating in
cosmic-ray air showers, there is an additional two-dimensional cut
in the plane of detected zenith angle, cos($\theta$), and log$_{10}$(NPE):

\begin{itemize}
  \item if cos($\theta$) $\le$ 0.1, then log$_{10}$(NPE) $>$ 3.6
  \item if cos($\theta$) $>$ 0.1, then log$_{10}$(NPE) $>$ 3.6 $+$ 2.99$\times\sqrt{1-(\frac{cos(\theta)-0.93}{0.83})^{2}}$
\end{itemize}

\begin{figure}[!ht]
\centering
\includegraphics[width=0.7\textwidth]{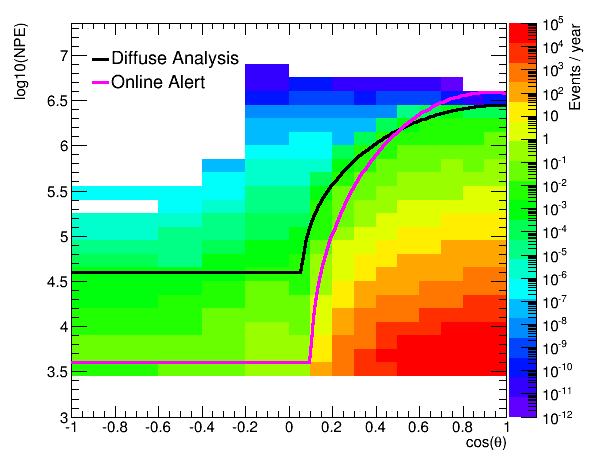}
\includegraphics[width=0.7\textwidth]{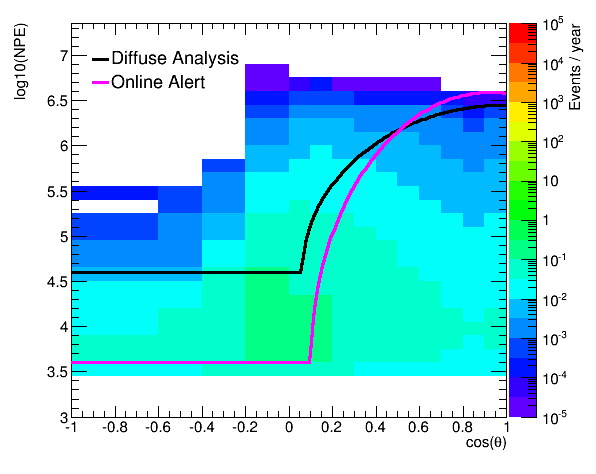}
\caption{The two-dimensional plane of cos($\theta$) and log$_{10}$(NPE) for
  the summed background (atmospheric muons and atmospheric neutrinos) on the upper panel and signal
  (astrophysical  $\nu_{\mu}$, assuming an E$^{-2}$ spectrum) on the lower panel.  Both the
  offline ultra-high-energy diffuse analysis (black~\cite{Aartsen:2013dsm}) and EHE Alert selection (magenta) are shown, with any
  event found above the line being selected.
  Few astrophysical signal events are expected for  cos($\theta$) $< \sim$-0.3
  at large log$_{10}$(NPE) values ($>\sim$~5.5) due to neutrino absorption in the Earth.}
\label{fig:sig_bkg_srs}
\end{figure}

\begin{figure}[!ht]
\centering
\includegraphics[width=0.6\textwidth]{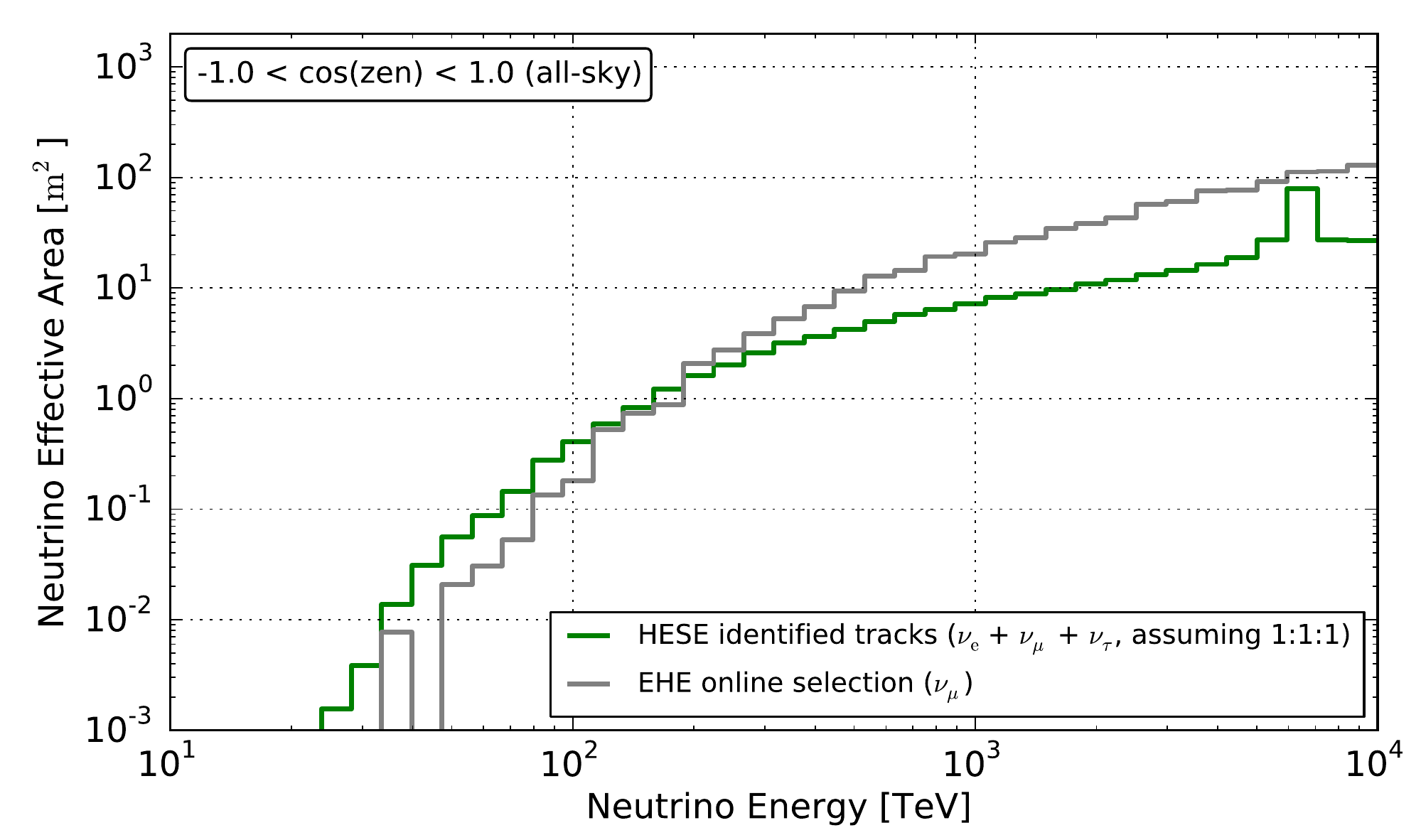}
\includegraphics[width=0.6\textwidth]{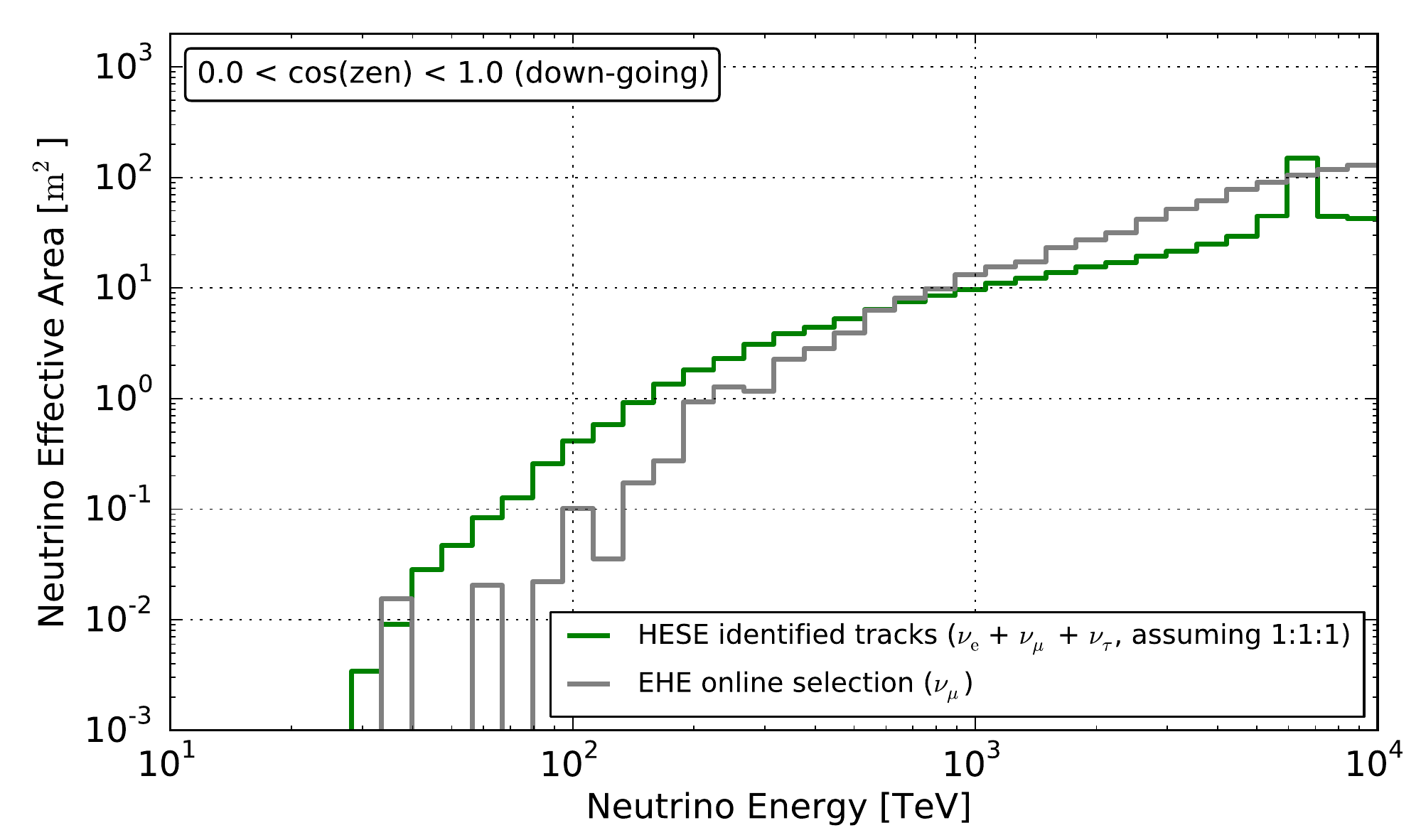}
\includegraphics[width=0.6\textwidth]{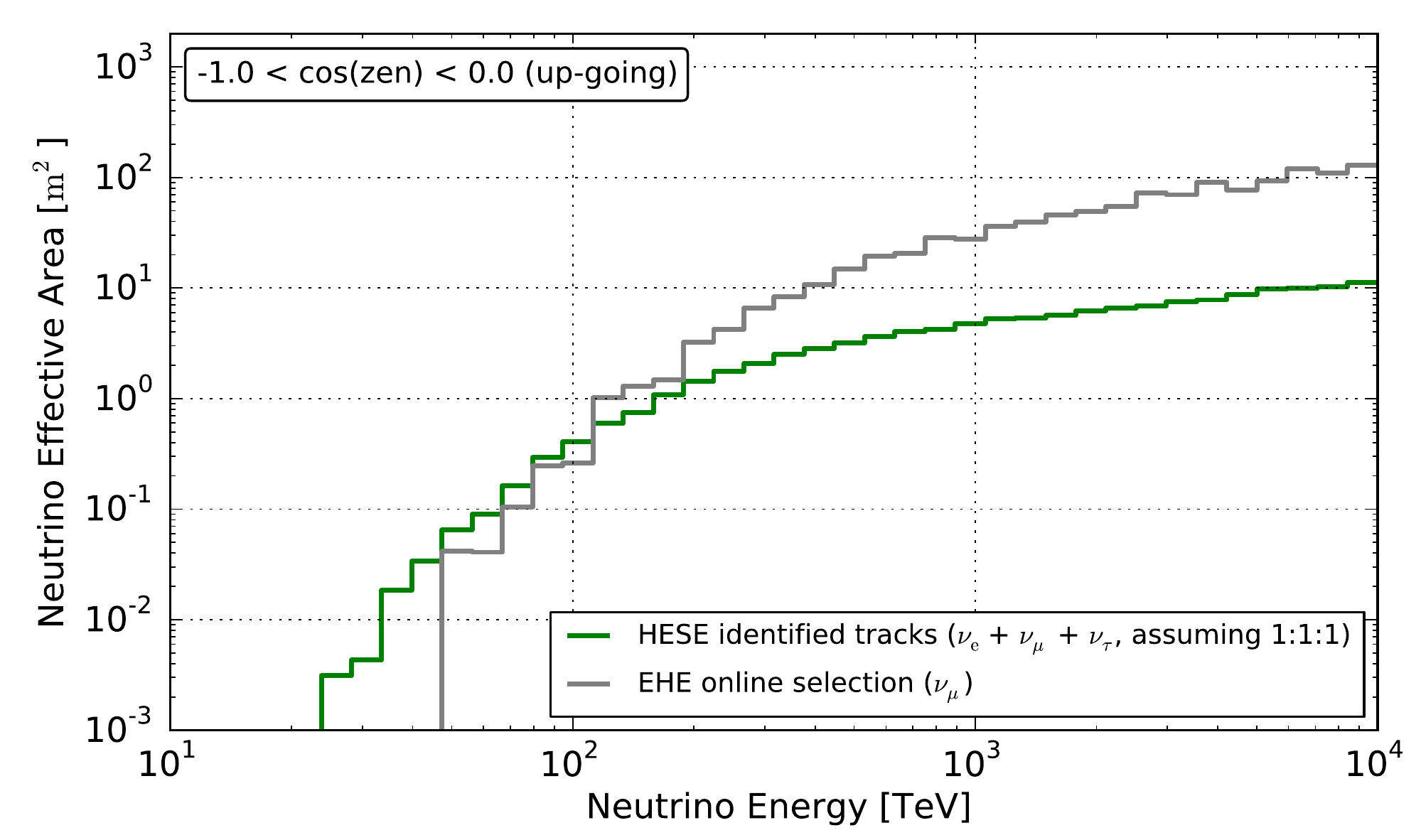}
\caption{The effective area as a function of neutrino energy for $\nu_{\mu}$ events for the online EHE signal diffuse neutrino selection
and the for the HESE identified tracks (presented as the sum of the three per-flavor
effective areas assuming 1:1:1 between neutrino flavors, and dominated by the $\nu_{\mu}$ component).
The top panel shows the all-sky effective area, while the down-going (southern sky)
and up-going (northern sky) effective areas are shown in the middle and bottom panels, respectively.  The EHE effective areas
are increased relative to the published offline selection~\cite{EHEPaper2}, while the HESE effective areas shown here are
just for track-like events and show reduced effective area relative to the published analysis~\cite{Aartsen:2014gkd}.}
\label{fig:ehe_effa}
\end{figure}

The two-dimensional selection was determined by optimizing for maximum signal retention assuming
an astrophysical E$^{-2}$ neutrino flux while tolerating some contamination from atmospheric backgrounds.
This requirement is illustrated in Fig.~\ref{fig:sig_bkg_srs} for the
total estimated background from simulation (atmospheric muons and atmospheric muon neutrinos) and for
a simulated astrophysical signal  assuming
$ \phi = 1.0\times10^{-18} (E/100 \,\rm{TeV})^{-2}$ [GeV$^{-1}$ cm$^{-2}$ s$^{-1}$ sr$^{-1}$]~\cite{Aartsen:2015rwa}.
Fig.~\ref{fig:sig_bkg_srs} also illustrates the increased signal acceptance for the online alert selection
compared to the offline diffuse analysis.  The neutrino effective area for the online selection is calculated
and is shown in Fig.~\ref{fig:ehe_effa} for the entire sky, as well as separately for the northern and southern hemispheres,
and reflects an overall increase in sensitivity to events in the TeV and PeV range with respect to the offline diffuse analysis.

The measured neutrino spectrum from contained event searches~\cite{ICRCKopper} and from through-going track
searches~\cite{Aartsen:2016xlq} return different spectral indices.  Given these uncertainties, the expected event rate for
signal is also estimated from simulated $\nu_{\mu}$ events weighted to a flux of
$ \phi = 2.3\times10^{-18} (E/100 \,\rm{TeV})^{-2.49}$ [GeV$^{-1}$ cm$^{-2}$ s$^{-1}$ sr$^{-1}$]~\cite{GlobalFit}
is also calculated.
The total number of alerts, classified by background type and signal assumption, is given in Table~\ref{tab:event_rate}.

\begin{table}[!ht]
\begin{center}
        \begin{tabular}{l | c}

        Sample & Events $/$ year \\
        \hline
        Atmospheric muon &                                  0.52  \\
        Conv. Atmos. $\nu_{\mu}$ &             1.20  \\
        Prompt Atmos. $\nu_{\mu}$ &             0.19 \\
        \hline
        Total Background &                              1.91 \\
        \hline
        Astro. $\nu_{\mu}$ (E$^{-2}$) &                             4.09 \\
        Astro. $\nu_{\mu}$ (E$^{-2.49}$) &                             2.48 \\

        \end{tabular}
        \caption {Summary of the expected event alert rate for the
          EHE online alert stream for each sample per year, including expected contributions from backgrounds and 2 astrophysical neutrino
          spectra.  The expected signal to noise ratio for this search is $\sim$2.}
        \label{tab:event_rate}
\end{center}
\end{table}

In addition to this prediction, the total alert rate has been validated
using four years of archival IceCube data.  The observed rate of 4.25 events per year is in agreement with
background + signal hypothesis if the IceCube astrophysical diffuse global fit spectral fit results~\cite{GlobalFit}
are assumed.    Each event found in the archival data search has been visually inspected
to confirm that they are, in fact, tracks.

\begin{figure}[!ht]
\centering
\includegraphics[width=0.7\textwidth]{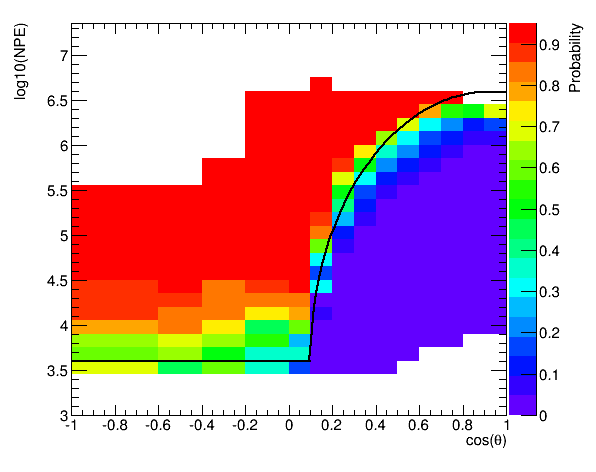}
\caption{The two-dimensional signalness map in the plane of
  cos($\theta$) and log$_{10}$(NPE) derived from simulated events.
  Bins are only filled if a minimum threshold of signal events
  is met, and empty bins are indicated by white space.}
\label{fig:ehe_probmap}
\end{figure}

A signalness parameter is calculated to provide a measure
of how likely each event is to be of astrophysical origin relative to the total background rate.
The signalness parameter is estimated from simulation by creating
a two-dimensional probability map in the plane of cos($\theta$) and log$_{10}$(NPE),
following
\begin{equation}
  {P_{i,j} = N^{sig}_{i,j} / (N^{sig}_{i,j} + N^{bkg}_{i,j})},
\end{equation}
where i and j are the bins in the 2D plane, N$^{sig}$ is the expected
astrophysical $\nu_{\mu}$ signal (assuming an E$^{-2}$ spectrum) in
each bin, and N$^{bkg}$ is the expected atmospheric muon and neutrino
background in the bin. The resulting map can be seen in
Fig.~\ref{fig:ehe_probmap}.  Bins are only filled if there is a minimum signal expectation
from simulations of at least 10$^{-4}$ events / year.  The probability is capped at 95\% due to limited
statistics available in each bin.  In the case that a data event
falls within the white space where we do not have Monte Carlo coverage,
the signalness value is set to -1, and no alert is sent.

Events passing the EHE alert selection will generate a public alert via the GCN network
sent via AMON.  These alerts
will contain the best-fit source direction and angular uncertainty from the online reconstruction, date and
time of the event, total measured charge, and measured signalness parameter.

\begin{figure}[ht]
\centering
        \includegraphics[width=0.7\textwidth]{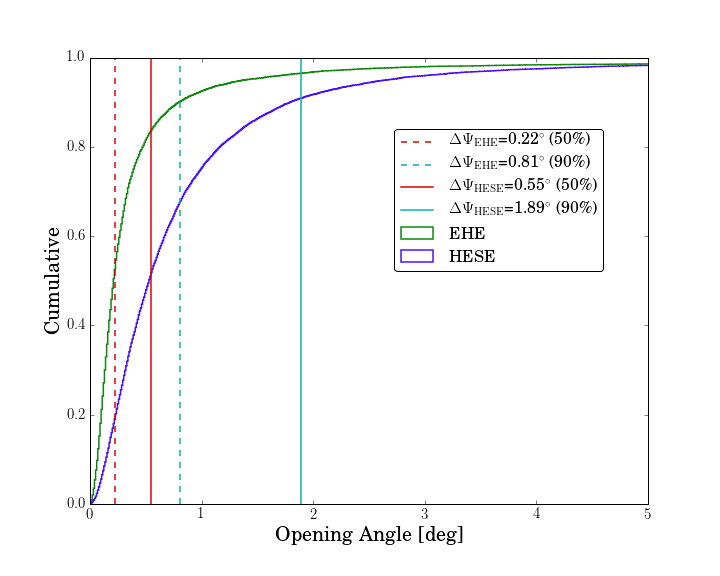}
        \caption{The cumulative distribution for the opening
          angle showing the radius containing 50\% and 90\% of
          simulated events for HESE and EHE selected neutrino events.
          The largest variation in the HESE angular resolution comes from reconstructed track length.
          Here ~80\% the HESE track-like events with a reconstructed track length $>$200~m are considered.
          The remaining ~20\% are reported with an upper limit fixed median angular error of $1.6^\circ$ ($8.9^\circ$ for 90\% containment).
          Systematic errors are not included.
        }
        \label{fig:angularres}
\end{figure}

The angular resolution has been studied utilizing simulated $\nu_{\mu}$ events
passing the EHE alert selection criteria by investigating
the opening angle between the best available online fit to the observed muon track
and the true neutrino direction.  The median angular
resolution is found to be 0.22$^{\circ}$, as illustrated in Fig.~\ref{fig:angularres},  and is nearly flat across the
neutrino energy spectrum.
As with the HESE alerts, after the initial alert generation, follow-up reconstructions of the EHE events by
computer clusters in the northern hemisphere are immediately started,
and revised directional coordinates and improved angular uncertainty are released within a
few hours of the initial alert (see $\S$\ref{sec:AlertsAndRevs}).

\section{Summary}
This paper provides a description of IceCube's realtime alert system, and the current active data alert streams.
The computing infrastructure at both the South Pole facility as well as in the
northern hemisphere allows IceCube to communicate, in real-time, the observation of
high quality candidate neutrino singlets and multiplets. Within this framework, several
analyses are currently implemented and sending alerts to our follow-up partners, as well as to the wider astronomical community
through AMON. This system enables
the existing analyses to be improved further and add new ones rapidly to respond to our evolving understanding of the
astrophysical neutrino signal observed by IceCube.

With the establishment of the IceCube realtime alert system, the alerts
generated by IceCube, and the potential discovery of transient astronomical sources in conjunction with them,
the era of multi-messenger time domain astronomy has arrived.  High-energy neutrinos are a unique messenger, able to travel astronomical
distances with negligible deflection or absorption, and clearly indicative of high energy hadrons in their sources.
A clear multi-messenger detection of a source holds the potential to enrich our understanding of the most energetic cosmic
phenomena, shed light on the mysterious origins of the highest energy cosmic rays,  and provide a unique window into the cosmos.


We acknowledge the support from the following agencies: U.S. National Science Foundation-Office of Polar Programs,
U.S. National Science Foundation- Physics Division, University of Wisconsin Alumni Research Foundation, the Grid Laboratory
Of Wisconsin (GLOW) grid infrastructure at the University of Wisconsin - Madison, the Open Science Grid (OSG) grid infrastructure;
U.S. Department of Energy, and National Energy Research Scientific Computing Center, the Louisiana Optical Network Initiative (LONI)
grid computing resources; Natural Sciences and Engineering Research Council of Canada, WestGrid and Compute/Calcul Canada;
Swedish Research Council, Swedish Polar Research Secretariat, Swedish National Infrastructure for Computing (SNIC), and Knut
and Alice Wallenberg Foundation, Sweden; German Ministry for Education and Research (BMBF), Deutsche Forschungsgemeinschaft (DFG),
Helmholtz Alliance for Astroparticle Physics (HAP), Research Department of Plasmas with Complex Interactions (Bochum), Germany;
Fund for Scientific Research (FNRS-FWO), FWO Odysseus programme, Flanders Institute to encourage scientific and technological
research in industry (IWT), Belgian Federal Science Policy Office (Belspo); University of Oxford, United Kingdom;
Marsden Fund, New Zealand; Australian Research Council; Japan Society for Promotion of Science (JSPS);
the Swiss National Science Foundation (SNSF), Switzerland; National Research Foundation of Korea (NRF); Villum Fonden,
Danish National Research Foundation (DNRF), Denmark

\bibliography{mybib}
\bibliographystyle{elsarticle-num}

\clearpage
\newpage


\end{document}